\documentstyle[aps,prl]{revtex}
\begin{document}
\draft
\author{Igor V. Barashenkov
\thanks{}
}
\address{Department of Applied Mathematics, University of Cape Town,\\
Private Bag Rondebosch 7700, South Africa}
\author{Alexander O. Harin
\thanks{}
}
\address{Department of Mathematics and Applied Mathematics,\\ University of
Natal, Durban 4001, South Africa}
\title{Nonrelativistic Chern-Simons theory for the Repulsive Bose
Gas\thanks{} }

\twocolumn[
\maketitle
\begin{minipage}{\textwidth}
\begin{quotation}
\begin{abstract}
We propose a  new nonrelativistic Chern-Simons theory
based on a simple modification of the standard Lagrangian.
This admits asymptotically
 nonvanishing field configurations
and is applicable to the description of systems of repulsive
bosons. The new model supports topological vortices and
 has a self-dual limit, both in the pure Chern-Simons and in the mixed
Chern-Simons-Maxwell cases. The analysis is based on a new formulation
of the Chern-Simons theories as constrained Hamiltonian
systems.
\bigskip

\bigskip
\end{abstract}
\end{quotation}
\end{minipage}]
\footnotetext{$^*$To appear in {\em Physical Review Letters}}
\footnotetext{$^{\dag}${\em E-mail\/}: igor@uctvax.uct.ac.za;
igor@maths.uct.ac.za}
\footnotetext{$^{\ddag}${\em E-mail\/}: harin@ph.und.ac.za;
harin@lourie.und.ac.za}

Chern-Simons vortices  are currently
considered as a candidate for anyon-like objects in quasiplanar
systems of condensed matter physics. Both  relativistic
and nonrelativistic matter fields can support these
structures~\cite{JP:prtph}. So far the
relativistic model seems to have exhibited a wider spectrum of solutions.
It was shown to possess both topological vortices, i.e.\ asymptotically
nonvanishing solutions with quantized flux~\cite{hong:lett,JW} and
nontopological solitons~\cite{JLW} for which    the matter field
$\psi$ vanishes at infinity. On the contrary, the {\em nonrelativistic\/}
model, which may happen to be more relevant for actual applications,
exhibits
only asymptotically vanishing solitons, or ``lumps"~\cite{JP}.
These structures arise when the interaction between the bosons is
attractive, $U \sim -(\psi \psi^*)^2$.
Since the lumps have nonzero vorticity and even some topological
characteristics (the flux is quantized),  they are sometimes
referred to as vortices and topological solitons. However, these are
clearly distinct
from what are usually called topological solitons on the plane, i.e.\
asymptotically nonvanishing solutions approaching different vacua in
different directions. The  ``genuine" topological vortices, or
``topological defects" interpolating between different nonzero vacua,
turned out to be absent in the nonrelativistic Chern-Simons theory,
though one could have expected them to arise in the case of
 {\em repulsive} nonlinearities $U(\psi)$.
Furthermore, it can be readily shown~\cite{BH} that  the nonzero
vacuum solution itself is not admitted by
the ``standard"  nonrelativistic model. There is no ``condensate"
in this theory, and it cannot be used to describe repulsive
boson gases.

The main objective of the present work is to incorporate the condensate
into the nonrelativistic gauge theory.
We show  that
the inapplicability of the standard model for the description of
asymptotically nonvanishing fields can be traced back
to an inadequate
Lagrangian formulation of its nongauged precursor. Appropriately modifying
this Lagrangian and using it as a basis for the gauge theory, we
arrive at a new version of the gauged nonlinear Schr\"odinger equation
which is completely compatible with the nonvanishing (``condensate")
boundary conditions.

We start with  the one-dimensional nonlinear Schr\"o\-dinger
 equation with a general nonlinearity,
\begin{equation}
i\psi_t+\psi_{xx}+F(\rho)\psi=0. \label{NLS}
\end{equation}
Here $\rho = |\psi|^2$, and $F(\rho) = -dU/d\rho$.
The standard Lagrangian for eq.~(\ref{NLS}),
\begin{equation}
\label{lagr:1dim}
{\cal L}=\frac{i}{2}(\psi_t \psi^*- \psi_t^*\psi)  - |\psi_x|^2  -U(\rho),
\end{equation}
does not {\em automatically\/} produce correct integrals of motion for
solutions
with  $|\psi|^2$ approaching $\rho_0$ at infinity.
First of all, the integral for the number of particles,
\begin{equation}
\label{fm:N}
N=\int\left(\frac{\partial{\cal L}}{\partial\psi_t}i\psi-
\frac{\partial{\cal L}}{\partial\psi^*_t}i\psi^*\right)dx,
\end{equation}
takes the form
$N=\int\rho\ dx$
 and apparently diverges. The regularized number of
particles,
\begin{equation}
\label{fm:mN}
N=\int(\rho-\rho_0)\, dx,
\end{equation}
is obtained by the {\it ad hoc\/} subtraction of the background
contribution.

The standard definition of momentum,
\begin{equation}
\label{fm:P}
P=\int\left(\psi_x\frac{\partial{\cal L}}{\partial\psi_t}+
\psi^*_x\frac{\partial {\cal L}}{\partial\psi^*_t}\right)dx,
\end{equation}
does not yield the correct expression either. For the
Lagrangian~(\ref{lagr:1dim}) one obtains
$P= (i/2)\int\left(\psi_x \psi^*-\psi^*_x\psi \right)dx$
which is  not compatible with  the Hamiltonian
structure of the model~\cite{bar:prepr}. Indeed, varying this $P$ gives
\[
\delta P=i\int(\psi_x\delta\psi^*-\psi^*_x\delta\psi)dx -
\rho_0\delta\text{Arg}\,\psi\big|_{-\infty}^{+\infty}.
\]
Due to the appearance of the boundary term we cannot compute
the
functional derivatives $\delta P/\delta\psi$ and $\delta P/\delta\psi^*$.
As a result, the Poisson bracket of $P$ with some other functional
(e.g.\ the Hamiltonian) can not be evaluated. To obtain the definition
compatible with the Hamiltonian structure  we again have to make
an {\em a posteriori\/} regularization~\cite{bar:prepr}
(see also~\cite{bog:FNT}):
\begin{eqnarray}
\nonumber
P&=&\frac{i}{2}\int\left(\psi^*\psi_x-\psi\psi^*_x\right)dx+
\rho_0\text{Arg}\,\psi\big|_{-\infty}^{+\infty}\\
\label{fm:mP}
&=&\frac{i}{2}\int(\psi^*\psi_x-\psi\psi^*_x)\left(1-
\frac{\rho_0}{\rho}\right)dx.
\end{eqnarray}

Proceeding to two dimensions, the standard definition
${\bf P}=(i/2)\int\left(
\psi^*\bbox{\nabla}\psi-\psi\bbox{\nabla}\psi^*\right)d^2r$
is even less suitable, since for asymptotically nonvanishing
configurations
this integral, in general, diverges.

Now we turn to the gauged nonlinear Schr\"odinger equation as
formulated by Jackiw and Pi~\cite{JP}:
\begin{mathletters}
\label{eq:nrfield}
\begin{equation}
i\psi_t -eA_0\psi  +{\bf D}^2\psi+F(\rho)\psi=0. \label{eq:gnls}
\end{equation}Here
$D_{\alpha} = \partial_{\alpha} + ieA_{\alpha}$ and the Abelian gauge field
$A_{\alpha}$ satisfies
\begin{equation}
\mu\partial_\beta F^{\beta\alpha}+
\frac{\kappa}{2}\epsilon^{\alpha\beta\gamma}F_{\beta\gamma}=e J^\alpha.
\label{eq:gf}
\end{equation}
\end{mathletters}Eq.~(\ref{eq:gf}) is the most general
linear gauge-field equation.
It combines both the Maxwell and Chern-Simons interactions.
$J^{\alpha}=(J_0, {\bf J})$ denotes the conserved matter current:
\begin{mathletters}
\begin{equation}
J_0=\rho=|\psi|^2,
\end{equation}
\vspace*{-\baselineskip}
\begin{equation}
{\bf J}=\frac{1}{i}\{\psi^*({\bf D}\psi)-\psi({\bf D}\psi)^*\}.
\label{J}
\end{equation}
\end{mathletters}$F_{\alpha\beta}
=\partial_{\alpha} A_{\beta} -  \partial_{\alpha} A_{\beta}$ and
Greek and Latin indices run over 0,1,2
and 1,2, respectively. The parameters
 $\mu$ and
$\kappa$
control the relative contributions of the Maxwell and Chern-Simons terms
in the corresponding
 Lagrangian:
\begin{eqnarray}
\nonumber
{\cal L}&=&\frac{i}{2}\left\{\psi^*(D_0\psi)-\psi(D_0\psi)^*\right\}-
(D_k\psi)^*(D_k\psi)\\
\label{lagr:2dim}
&&-\frac{\mu}{4}F_{\alpha\beta}F^{\alpha\beta}+
\frac{\kappa}{4}\epsilon^{\gamma\alpha\beta}A_\gamma F_{\alpha\beta}-
U(\rho).
\end{eqnarray}
$\mu$ is $\geq0$; we also assume  $\kappa\geq0$.
The case of negative $\kappa$ is recovered  by
the parity transformation ($x^1 \rightleftharpoons x^2$,
$A^1 \rightleftharpoons A^2$).

This system inherits the drawback of its nongauged counterpart.
 The momentum, defined by
\begin{equation}
P_i=\int\left[\frac{\partial\psi}{\partial x^i}\frac{\partial{\cal L}}
{\partial\psi_t}+
\frac{\partial\psi^*}{\partial x^i}\frac{\partial{\cal
L}}{\partial\psi^*_t}+
\frac{\partial A^{\alpha}}{\partial x^i}\frac{\partial{\cal L}}{\partial
A^{\alpha}_t}\right]\,d^2x,
\end{equation}
equals
\begin{eqnarray}
\nonumber
P_i&=&\int\left[\frac{i}{2}\left\{\psi^*(D_i\psi)-\psi(D_i\psi)^*\right\}-
\mu\epsilon_{ij}E^j B\right]d^2x\\
&=&\int\rho\,(eA^i-\partial_i\text{Arg}\,\psi)\,d^2x-
\mu\int\epsilon_{ij}E^j B\,d^2x.
\end{eqnarray}
Here $E^i=F^{i0}$ and $B=F^{21}$ are electric and magnetic
fields, respectively. The momentum in this case  is finite provided
$A^i \rightarrow(1/e)\partial_i\text{Arg}\,\psi$  at infinity,
but the functional derivatives $\delta P_i/\delta\psi$ and
$\delta P_i/\delta\psi^*$ still cannot be calculated.

Returning to the one-dimensional nongauged {\sc nls}~(\ref{NLS}), a
 natural question is: Is it possible to find a Lagrangian
 producing the same Euler-Lagrange equations
as eq.~(\ref{lagr:1dim}) but at the same time {\em automatically\/}
yielding
the correct integrals of motion? The answer is yes. The Lagrangian
\begin{equation}
\label{lagr:1dimmd}
{\cal L}=\frac{i}{2}(\psi_t \psi^*-
\psi_t^*\psi)\left(1-\frac{\rho_0}{\rho}\right)- |\psi_x|^2-U(\rho),
\end{equation}
produces the same {\sc nls}~(\ref{NLS}),
while formulas~(\ref{fm:N}) and~(\ref{fm:P}) yield the regularized
integrals
$N$ and $P$, as given exactly by~(\ref{fm:mN}) and~(\ref{fm:mP}).

One may argue that the modification is in a sense trivial, since
eqs.~(\ref{lagr:1dimmd}) and (\ref{lagr:1dim}) differ only by a total
derivative.
However, this is no longer the case if the gauge field is added:
\begin{eqnarray}
\nonumber
{\cal L}&=&\frac{i}{2}\left\{\psi^*(D_0\psi)-\psi(D_0\psi)^*\right\}
\left(1-\frac{\rho_0}{\rho}\right)-(D_k\psi)^*(D_k\psi)\\
\label{lagr:2dimmd}
&&-\frac{\mu}{4}F_{\alpha\beta}F^{\alpha\beta}+
\frac{\kappa}{4}\epsilon^{\gamma\alpha\beta}A_\gamma F_{\alpha\beta}-
U(\rho).
\end{eqnarray}
This is a {\em new\/} system. The field equations resulting
from~(\ref{lagr:2dimmd})
differ from those of the standard model~(\ref{lagr:2dim}). The difference
 lies in the definition of the number density:
\begin{equation}
J_0=\rho-\rho_0.
\label{J0}
\end{equation}
Apart from that,
the field equations are the same, eqs.~(\ref{eq:nrfield}),
with the spatial part of the current being given by the standard
expression (\ref{J}).
However,
the new system does possess the condensate solution, while the
 Lagrangian~(\ref{lagr:2dimmd}) produces the correct number of
particles, $N=\int(\rho-\rho_0)\,d^2r$,
and the momentum which is compatible with the Hamiltonian
structure of the model:
\begin{eqnarray}
\nonumber
P_i&=&\int\left[\frac{i}{2}\left\{\psi^* (D_i\psi)-\psi(D_i\psi)^*\right\}
\left(1-\frac{\rho_0}{\rho}\right)\right.\\
\nonumber
&&-\mu\epsilon_{ij}E^j B\bigg]d^2x\\
&=&\int\left[(\rho-\rho_0)
\left(eA^i-\partial_i\text{Arg}\,\psi\right)-
\mu\epsilon_{ij}E^j B\right]d^2x.
\end{eqnarray}
Physically, eq.~(\ref{J0}) implies that the condensate is electrically
neutral. This may result for instance from the presence of motionless
particles of the opposite charge.

The Hamiltonian formulation of this model deserves a special comment.
For the nonvanishing Maxwell term ($\mu \neq 0$), it can be
formulated only as a {\em constrained} \/ Hamiltonian system.
The Hamiltonian reads
\begin{equation}
\label{eq:energy}
H=\int \left[ |{\bf D}\psi|^2+U(\rho)
 +\frac{\mu}{2}\,({\bf E}^2+B^2)   \right]d^2x
\end{equation}
and the constraint is the time component of eq.~(\ref{eq:gf})
(the modified Gauss law):
\begin{equation}
\label{Gauss}
\mu \,\text{div}\,{{\bf E}} - \kappa B = e(\rho-\rho_0).
\end{equation}
The first pair of canonically conjugate variables is $\psi$ and $\psi^*$,
as usual. The other two canonical coordinates are $A^1$ and $A^2$, with
the corresponding conjugate momenta being
\begin{equation}
\label{conj:mom}
\Pi_i = \frac{\partial {\cal L}}{\partial (\partial_0
A^i)}=\frac{\kappa}{2}
\epsilon_{ij} A^j - \mu F^{i0}.
\end{equation}
Eqs.~(\ref{eq:nrfield}) can be represented then as
\[
i\psi_t = \frac{\delta \tilde H}{\delta \psi^*}, \quad
i\psi^*_t = - \frac{\delta \tilde H}{\delta \psi}, \quad
A^i_t = \frac{\delta \tilde H}{\delta \Pi_i}, \quad
{\Pi_i}_t = -\frac{\delta \tilde H}{\delta A^i},
\]
where $\tilde H$ is the Lagrange function,
\begin{equation}
{\tilde H} = H + \int A_0 \left\{ e(\rho - \rho_0)-
 \mu \,\text{div}{\bf E} +
\kappa B\right\} d^2x,
\nonumber
\end{equation}
and $A_0 = A_0({\bf x},t)$ plays
 the r\^ole of the Lagrange multiplier.
This formulation remains valid, of course, for the standard model
($\rho_0=0$.) In  the pure Chern-Simons case ($\mu=0$)
the constraint (\ref{Gauss}) can be  explicitly resolved which produces
an alternative, unconstrained formulation~\cite{JP}.

The condensate solution has the form:
$\psi=\sqrt\rho_0$, $A_0=(1/e)F(\rho_0)$, ${\bf A}=0$.
This solution
exists even when the potential $U(\rho)$ does not possess a symmetry
breaking
minimum at $\rho_0$. We can however confine ourselves to potentials with
$U'(\rho_0)=0$ as this condition can always be accomplished by the
transformation $A_0\rightarrow A_0+F(\rho_0)/e$,
$U(\rho)\rightarrow U(\rho)-F(\rho_0)\rho$.  The condensate solution
is then
$\psi=\sqrt\rho_0$, $A_0=0$, ${\bf A}=0$, and it
corresponds to  an extremum of $U(\rho)$.
Through the gauge invariance, the condensate generates
a set of singular solutions,
\begin{equation}
\label{background}
\psi=\sqrt\rho_0 e^{in\theta}, \quad A_0=0, \quad
{\bf A}=\frac{n}{e}\frac{{\bf e}_\theta}{r}
\end{equation}
which serve as  asymptotes for vortices as $r \rightarrow \infty$.
The magnetic flux of the vortex is quantized:
$\int B\, d^2x =\oint A_{\theta}\,rd\theta=
(2\pi/e)n$. Integrating  eq.~(\ref{Gauss})
over the entire plane we observe that the flux and the
charge of the vortex $Q=e\int (\rho-\rho_0)\,d^2x$ are related:
$-\kappa \int B\,d^2x=Q$.
Hence, $Q$ is quantized as well:
\begin{equation}
\label{eq:Nrho}
Q=e\int (\rho-\rho_0)\, d^2x=-\frac{2\pi\kappa}{e} n.
\end{equation}

Confining consideration to  radially symmetric configurations, we write
\begin{mathletters}
\begin{eqnarray}
&&\psi({\bf x})=\psi(r) e^{in\theta}, \\
&&A_0({\bf x})=A_0(r),\\
\label{eq:Aj}
&&A^j({\bf x})=\epsilon^{jk}x_k\frac{\Phi(r)}{r^2}, \quad j,k=1,2.
\end{eqnarray}
\end{mathletters}The magnetic field $B(r) = (1/r) \Phi_r$. Regularity
at the origin requires $\psi(0)=0$ when $n \neq 0$.
For finite
${\bf A}(0)$ we should also have $\Phi(0)=0$. Now making the gauge
transformation $\text{Arg}\, \psi \rightarrow \text{Arg}\, \psi -
n\,\theta$,
$\quad \Phi\rightarrow \Phi-n/e$,
the system~(\ref{eq:nrfield}) is cast in the form
\begin{mathletters}
\label{eqs:radsym}
\begin{eqnarray}
&&\Delta\psi-e^2\frac{\Phi^2}{r^2}\psi-e A_0\psi+F(\psi^2)\psi=0,\\
&&\mu B_r +\kappa \frac{dA_0}{dr}
-2e^2\frac{\psi^2}{r} \Phi=0,\\
&&\mu\Delta A_0+\kappa B+e(\psi^2-\rho_0)=0.
\end{eqnarray}
\end{mathletters}This
system does not contain $n$ explicitly. Vortices
with different vorticities will be distinguished
by the  boundary condition at the origin: $\Phi(0) =-n/e$.
Eq.~(\ref{eq:Nrho}) indicates that conventionally shaped topological
vortices
 for which $\rho(r)$
grows monotonically from $0$ to $\rho_0$,
are of positive vorticity. Vorticity-free solutions and those with $n<0$
have to
   be nodal, i.e.\ $\rho(r)-\rho_0$ should
necessarily change
sign.
Relegating the comprehensive discussion of solutions to the
system~(\ref{eqs:radsym})
to a more detailed publication~\cite{bh:prep}, here we consider
only an important special case.

Static solutions correspond to stationary points of the
Hamiltonian (\ref{eq:energy}) on the constraint manifold (\ref{Gauss}).
In the mixed CS-Maxwell case, using the flux-vorticity relation,
$\int B\, d^2x = (2\pi/e)n$
and the Bogomol'nyi decomposition,
\begin{equation}
\label{eq:decomp}
|{\bf D}\psi|^2=|(D_1\pm i D_2)\psi|^2
\pm\frac{1}{2}\bbox{\nabla}\times{\bf J}\pm e B \rho,
\end{equation}
the energy (\ref{eq:energy}) takes the form
\begin{eqnarray}
\nonumber
H&=&\int \left\{|(D_1\pm i D_2)\psi|^2+
\frac{\mu}{2}\left[B\pm\frac{e}{\mu}(\rho-\rho_0)\right]^2\right.\\
&&-\left.\frac{e^2}{2\mu}(\rho-\rho_0)^2+U(\rho) +
\frac{\mu}{2}(\bbox{\nabla} A_0)^2
\pm\frac{1}{2}\bbox{\nabla}\times{\bf J}\right\}d^2x\nonumber\\
&&\pm2\pi\rho_0n.
\end{eqnarray}
For $U(\rho)=(e^2/2 \mu)(\rho-\rho_0)^2$
(which corresponds to the Bose gas with the $\delta$-function pairwise
repulsion)
and fields approaching
the condensate background (\ref{background}), the energy can be rewritten
as
\begin{eqnarray}
H&=&\int\left\{|(D_1\pm i D_2)\psi|^2+
\frac{\mu}{2}\left[ B\pm \frac{e}{\mu}(\rho-\rho_0)\right]^2\right.\\
\nonumber
&&\left.+\frac{\mu}{2}(\bbox{\nabla} A_0)^2\right\} d^2x
\pm 2\pi\rho_0 n.
\end{eqnarray}
The lower bound of energy, $H=\pm 2\pi\rho_0 n$, is saturated when the
following self-duality equations are satisfied:
\begin{mathletters}
\label{eqs:selfdual}
\begin{eqnarray}
&&(D_1\pm i D_2)\psi=0,\label{sd:1}\\
&&B\pm\frac{e}{\mu}(\rho-\rho_0)=0,\label{sd:2}\\
&&A_0=0.
\end{eqnarray}
\end{mathletters}The upper (lower) sign should be associated
with the positive (negative) vorticity $n$. This can be concluded,
for example, simply  from the fact that the energy is positive for
positive $U(\rho)$, see eq.~(\ref{eq:energy}).
Comparing (\ref{sd:2})
to (\ref{Gauss}), we see that solutions of eqs.~(\ref{eqs:selfdual})
lie on the constraint manifold only for $\kappa =\mu$ and only in the case
of the upper sign. Thus, for $\kappa >0$, only
 vortices with positive vorticity may exist.
Eq.~(\ref{sd:1}) yields
\begin{equation}
A^i = \pm \frac{1}{2e}\epsilon_{ij}
\partial_j \ln{\rho} + \frac{1}{e}\partial_i {\rm Arg} \psi,
\label{A}
\end{equation}
and we should retain only the upper sign here.
Substituting this into (\ref{sd:2}), we arrive at
\begin{equation}
\label{eq:ln}
\nabla^2\ln\rho=2\frac{e^2}{\kappa} (\rho-\rho_0).
\end{equation}
Eq.~(\ref{eq:ln}) appeared  previously in  the self-dual limit
of   the relativistic Higgs model with Maxwell term~\cite{bogom:sjnf} and
is known to possess  solutions
with  ``topological vortex" asymptotic behaviour:
$\rho(\infty)=\rho_0$, $\rho(0)\sim r^{2n}$,
$n\geq 1$. The phase space analysis shows that it also admits
lump solitons vanishing at infinity~\cite{bh:prep},
 but these are extraneous to the
subject of the present work.

The self-duality reduction is also possible  in the pure Chern-Simons case
($\mu=0$). Making use of the identity~(\ref{eq:decomp}) and the
constraint~(\ref{Gauss}),
the energy~(\ref{eq:energy}) can be represented as
\[
H=\int\left[|(D_1\pm i
D_2)\psi|^2\mp\frac{e^2}{\kappa}(\rho-\rho_0)\rho+U(\rho)\right] d^2x.
\]
Invoking  eq.~(\ref{eq:Nrho}),
 we rewrite this as
\begin{eqnarray}
\nonumber
H&=&\int\left[|(D_1\pm i D_2)\psi|^2\mp
\frac{e^2}{\kappa}(\rho-\rho_0)^2+U(\rho)\right]d^2x\\
&&\pm2\pi\rho_0n.
\end{eqnarray}
With the choice of
$U(\rho)=\pm(e^2/\kappa)(\rho-\rho_0)^2$ one observes that
the energy is minimal provided $(D_1 \pm iD_2)\psi =0$,
whence we have, as before, eq.~(\ref{A}). Substituting this into
the constraint equation (\ref{Gauss}), we arrive at
\begin{equation}
\label{eq:lnpm}
\nabla^2\ln\rho=\pm2\frac{e^2}{\kappa}(\rho-\rho_0).
\end{equation}
Note that no equation for the scalar potential $A_0$ arises here.
This is not surprising as $A_0$ is not a dynamical variable.
(In our Hamiltonian formulation, it is just a Lagrange multiplier.)
For any combination of the Maxwell and Chern-Simons terms
$A_0$ can be determined from the spatial part of
 eq.~(\ref{eq:gf})~\cite{comment}:
\begin{equation}
\mu \bbox{\nabla} \times B + \kappa \bbox{\nabla} \times A_0 = e {\bf J}.
\label{spatial}
\end{equation}
The matter current is calculated from eq.~(\ref{A}):
 ${\bf J} = \mp \bbox{\nabla} \times
\rho$. Substituting this into eq.~(\ref{spatial}) for $\mu =0$,
we can readily solve for $A_0$:\vspace{-0.25cm}
\begin{equation}
A_0 = \mp \frac{e}{\kappa} (\rho -\rho_0).
\label{A_0}
\end{equation}

In contrast to the mixed CS-Maxwell case, {\em both\/} signs are
allowed in (\ref{eq:lnpm}) and so
the pure Chern-Simons model exhibits  a wider variety of self-dual
solutions \cite{bh:prep}. Here we only dwell on the
repulsive case,
$U = (e^2/\kappa)(\rho-\rho_0)^2$
(positive sign in (\ref{eq:lnpm})).
Our remark below pertains to the structure of the nonrelativistic Chern-
Simons
vortices. As we have already mentioned, the ``vortex" solutions
of eq.~(\ref{eq:ln}) are conventionally shaped:
$\rho(r)$ is $\sim r^{2n}$ near the origin, and exponentially approaches
$\rho_0$ as $r \rightarrow \infty$. Eq.~(\ref{Gauss})
 with $\mu=0$ implies that the magnetic field is greatest
at the centre of the vortex, and rapidly decays  to zero
as $r \! \rightarrow \!\infty$. (This is in contrast to the relativistic
case \cite{hong:lett,JW} where the magnetic field is concentrated
in a ring.) Eq.~(\ref{A_0}) meanwhile indicates that the electric field,
${\bf E}=-{\bf e_r}dA_0/dr$, is localized within a ring around the
centre of the vortex and vanishes both at $r\!=\!0$ and $r\! \rightarrow
\! \infty$.
The structure of the mixed CS-Maxwell self-dual vortex is the
same, with the exception that it carries no electric field.

We thank Professors P.~G.~L.~Leach and P.~Wiegmann
for  their useful remarks
about this work.
This research was  supported by the Foundation for Research Development
of South Africa and the Visiting Lecturer's Fund of UCT.


\begin{thebibliography}{11}
\bibitem{comment}
In the pure Chern-Simons case  eq.~(\ref{Gauss}) can be solved
for {\bf A} and so the vector potential can also be treated as a dependent
variable. This is the formulation utilized by Jackiw and Pi~\cite{JP},
whereas we treat  {\bf A} as a dynamical variable.
The resulting equations are, of course, equivalent.
\end{thebibliography}
\end{document}